\documentclass[12pt,preprint]{aastex}

\begin{document}
\def\lsun{{\rm L_{\odot}}}
\def\msun{{\rm M_{\odot}}}
\def\rsun{{\rm R_{\odot}}}
\def\go{
\mathrel{\raise.3ex\hbox{$>$}\mkern-14mu\lower0.6ex\hbox{$\sim$}}
}
\def\lo{
\mathrel{\raise.3ex\hbox{$<$}\mkern-14mu\lower0.6ex\hbox{$\sim$}}
}
\def\simeq{
\mathrel{\raise.3ex\hbox{$\sim$}\mkern-14mu\lower0.4ex\hbox{$-$}}
}

\input epsf.sty

\title{The Stars of the Galactic Center} 
\author{Melvyn B.\ Davies$^1$, and Andrew King$^2$\\ 
1. Lund Observatory, Box 43, SE--221 00 Lund, Sweden; mbd@astro.lu.se \\
2. Theoretical Astrophysics Group, University of Leicester, Leicester LE1
7RH, UK; ark@astro.le.ac.uk}

\begin{abstract}
We consider the origin of the so--called S~stars orbiting the
supermassive black hole at the very center of the Galaxy. These are
usually assumed to be massive main--sequence stars. We argue instead
that they are the remnants of low--to--intermediate mass red giants
which have been scattered on to near--radial orbits and tidally
stripped as they approach the central black hole. Such stars retain
only low--mass envelopes and thus have high effective
temperatures. Our picture simultaneously explains why S~stars have
tightly--bound orbits, and the observed depletion of red giants in the
very center of the Galaxy.
\end{abstract}

\section{Introduction}

Proper motion observations (Eckert \& Genzel 1997, Ghez et al 1998)
have revealed in exquisite detail the motions of stars around the
black hole in the center of the Galaxy. The closest known stars have
extremely eccentric orbits (for example S2 has an eccentricity of
0.87), pericenters of only a few hundred AU, and orbital periods in
the range 15 -- 300~yr (Sch\"odel et al 2003, Eisenhauer et al 2005).
These ``S~stars'' are all rather blue ($T_{\rm eff} \sim 30000$~K.
These properties lead to the ``birth problem'' (Ghez et al 2003). The
S~stars orbit in a region where tidal forces from the central
supermassive black hole prevent star formation {\it in situ},
although several authors have examined whether they were
produced after a collision of dense molecular clouds
or in an AGN--like disc (e.g. Genzel et al. 2003; Levin
\& Beloborodov 2003; Milo\v savljevi\'{c} \& Loeb 2004). A
plausible alternative picture is that the S~stars result from the
sinking of massive stellar clusters towards the black hole by
dynamical friction. However tides disrupt such clusters at distances
$\sim $ 1 pc from the black hole, much further out than the region of
the observed S stars (Gerhard 2001). To supply the S~stars requires
scattering into near--radial orbits by gravitational interactions with
other stars. However this process occurs on the local relaxation
timescale, which would considerably exceed the main--sequence
lifetime (around 10 Myr) of stars of the observed temperatures.

One way out of this problem is to increase the stability of the
clusters against tides. There have been several suggestions that
stellar clusters could contain intermediate--mass black holes, making
them less vulnerable to tidal shredding (Hansen \& 
Milo\v savljevi\'{c}
2003, McMillan \& Portegies Zwart 2003, Kim et al 2004). We do not
adopt this idea in this {\it Letter}.

These closest stars have another intriguing property: there appear to
be no red giants among them, despite the fact that these would be
brighter than main--sequence stars in the infrared K band where the
observations are made. We call this the red--giant depletion problem.
It has been suggested that such stars are destroyed in stellar
collisions (e.g. Genzel et al. 1996; Alexander 1999) although
collisions have been shown to be inefficient at removing most
of the red--giant envelope (Bailey \& Davies 1999).

A third problem arises from a selection effect for identifying stars
moving very close to the black hole pointed out by Sch\"odel et al
(2003). Namely, stars with pericenters similar to the S~stars but much
longer orbital periods would spend almost all of their lifetimes far
from pericenter moving very slowly, and thus not turn up in proper
motion observations. In other words, only tightly bound orbits are
easily identifiable, and a valid picture of the origin of the S~stars
must explain these orbits too. We call this the orbit problem. 

In this {\it Letter} we suggest a simultaneous solution to the birth,
red--giant depletion and orbit problems.

\section{Violent Mass Loss}

Table 1 gives the known pericenter distances $p$ and
eccentricities for S~stars. The
instantaneous tidal lobe of a star at pericenter has size

\begin{equation}
R_T \simeq 0.5\biggl({M_*\over M_{\rm BH}}\biggr)^{1/3}p  \label{lobe}
\end{equation}
where $M_*, M_{\rm BH}$ are the masses of the star and central black
hole.  Table 1 gives the values of $R_T$ corresponding to the observed
S star orbits, taking values of $3\msun$ and $3.6\times 10^6\msun$ for
$M_*, M_{\rm BH}$. In four cases (including those with the
most accurate orbital determinations) we note that these are of
order a few $100\rsun$.  This is comparable in size to a
low--to--moderate mass star on the AGB, with core masses $\sim 0.6 -
0.8$ M$_\odot$. This can be seen in Fig. 1, where we plot the
evolutionary tracks for solar metallicity stars of masses 1, 2, 3, and
4 M$_\odot$ using the {\it SSE} code of Hurley, Pols \& Tout (2000).

We can now ask what would happen to an AGB star scattered on to an
orbit passing close enough to the black hole that its radius exceeds
its tidal radius.  Tidal disruption of stars in the center of our own
and other galaxies has been discussed in some detail (e.g. Alexander
\& Livio 2001; Di Stefano et al 2001; Ghez et al 2003).  At its first
pericenter passage, the scattered star loses most of its envelope in a
violent fashion, leaving only the carbon--oxygen core, the
shell--burning layers, and a tenuous envelope.  Observations of the
secondary star of the X--ray binary Cyg X--2 (Casares et al., 1998;
Orosz \& Kuulkers 1999) show that it still has hydrogen in its
envelope, and that it has a luminosity $ L \simeq 150$ L$_\odot$
despite a mass of only $\simeq$ 0.8M$_\odot$. These properties agree
with binary evolution models where the secondary descends from a 3.5
M$_\odot$ star that has lost most of its mass quite rapidly (King \&
Ritter, 1999, Kolb et al, 2000, Podsiadlowski et al 2002).  These
models suggest that the envelope mass is less than 0.1 M$_\odot$ on
top of a 0.7 M$_\odot$ core, but the star closely resembles a 3.5
M$_\odot$ main--sequence star with $ L \simeq 150$ L$_\odot$. We
expect that an AGB star that encounters the black hole will be left
with a similar envelope mass.  The star retains the same nuclear
luminosity but is much smaller and bluer.  Stripped AGB stars have
luminosities around $10^4$ L$_\odot$ and temperatures $\ga 10^4$~K,
similar to B0 main--sequence stars. This is consistent with the idea
that the S~stars were all red giants which were violently stripped of
their envelopes near pericenter. The nuclear lifetimes of these
stripped stars would be a few Myr, since they have $\la 0.1\msun$ of
envelope available for nuclear burning (see the Discussion). Stars
must be captured and stripped at a rate $\sim 10^{-6}~{\rm yr}^{-1}$
to prevent the current population of S~stars decaying on this
timescale. Since the tidal interaction is effectively pointlike, the
pericenters of the stripped stars remain unchanged by it. As seen from
the Table, this kind of evolution is possible for the current orbits
of at least four of the S~stars. We discuss the other two stars later
(Section 3).

\begin{deluxetable}{lrrrrrr}
%\tabletypesize{\scriptsize}
\tablecaption{Tidal radii for measured orbits.}  \tablewidth{0pt}
\tablehead{ \colhead{Name} & \colhead{$p$ (mpc)} & 
\colhead{$e$} &  \colhead{$R_T$
(R$_\odot$)} & \colhead{$p$ (mpc)} & \colhead{$e$} &
\colhead{$R_T$ (R$_\odot$)} }
\startdata 
S2 & 0.6 & 0.87 &114 & 0.6 & 0.88 &111 \\
 S12 & 2.2 & 0.73 &424 & 1.1 & 0.90 &206 \\
 S14 & 0.4 & 0.97 & 75 & 0.5 & 0.94 &101 \\ 
S1 & 8.0 & 0.62 &1541& 10.1 & 0.36 &1941 \\ 
S8 & 0.6 & 0.98 & 116& 0.9 & 0.93 &176 \\ 
S13 & 6.0 & 0.47 &1156& 5.0 & 0.40 &963 \\ 
\enddata
\tablecomments{The 2nd--4th columns are from the orbital fits of
Sch\"odel et al (2003), while the 5th--7th columns are from the
fits of Eisenhauer et al (2005). Note that the orbits for S14, S1, S8,
and S13 were all poorly constrained in Sch\"odel et al.}
\end{deluxetable}

Stripping stars solves the birth problem: if the S~stars are not
main--sequence stars with masses $\sim 10 - 15 \msun$, but instead
stripped giants of orginal (main--sequence) masses $\sim 1-4 \msun$,
their ages are greater than 300 -- 3000 Myr, allowing ample time for
them to be scattered from star clusters further out into orbits with
the current small pericenter distances. The violent tidal mass loss we
have invoked leaves the pericenter distances unchanged, but produces
the observed tightly bound orbits, as orbital energy is used to remove
the stellar envelope.

To see this we compute the energy loss required to take a star from a
near--parabolic orbit to a tightly bound one like those seen,
and compare it to the energy needed to strip a red giant down to a
core of mass $M_c$ and radius $R_c$. The loss of
orbital binding energy is
\begin{equation}
E_{\rm orb} = {GM_{\rm BH}M_c\over 2a}, \label{binding}
\end{equation}
%while the gravitational binding energy of the red--giant envelope is
%\begin{equation}
%E_{\rm env} = {{GM_\star}M_e\over \lambda R_\star} \label{envelope}
%\end{equation}
%where for example $\lambda \sim 0.5$ for a 3M$_\odot$ AGB with a
%radius $R_\star \simeq 300 \rsun$ (Dewi \& Tauris 2000). Thus
%\begin{equation}
%{E_{\rm orb} \over E_{\rm env}} = {M_{\rm BH}M_c \over M_\star M_e}
%{R_\star \over 4 a} \gg 1
%\end{equation}
%for values of $a \sim 40$ mpc. We see that the bulk of the red giant envelope
%is easily ejected. We are, however, interested in removing essentially {\it all}
%the envelope, leaving only the core and the shell--burning layers, otherwise the red
%giant would be rejuvenated if more than about 0.05 M$_\odot$ of the envelope remains.
%We therefore need to dig deeply into the envelope removing the gas at some core
%radius, $R_c$, from the center of the star.
where $a$ is the semi--major axis after the mass loss. To strip the
star down to the core plus a burning shell and a very tenuous envelope
requires an energy
\begin{equation}
E_{\rm strip} \sim {GM_cM_e\over R_c} \label{strip}
\end{equation}
where $M_e$ is the initial envelope mass. Thus stripping requires
\begin{equation}
a\la 0.5{M_{\rm BH}\over M_e}R_c \simeq 2\times 10^{16}\left({M_e\over 
\msun}\right)^{-1} {R_c\over 10^{10}~{\rm cm}}~{\rm cm}  \label{separation}
\end{equation}
The resulting values of $a$ compare well with those measured for the
S~stars listed in Table 1, and imply periods in the observed range
$\la 100$~yr. The observed pericenters $p = a(1-e)$ require eccentricities
\begin{equation}
e \simeq 1 - 2{pM_e\over R_cM_{\rm BH}} \simeq 0.9
\end{equation}

The idea of violent tidal mass loss thus explains the current tight
orbits of most of the S~stars as well as solving the birth problem. We
can extend this to a simultaneous solution of the red--giant depletion
problem by noting our conclusion at the end of the Introduction above
that only tightly--bound orbits are easily identifiable. Red giants on
orbits such that they do not fill their tidal lobes near pericenter
have periods of many centuries and are very unlikely to be identified
as high proper motion objects. Hence the only stars potentially
identifiable as Galactic Center red giants lose their envelopes and
turn into S~stars instead.

\section{Scattering of Stars}

We noted above that two of the S~stars (S1 and S13) would not have
filled their tidal lobes as red giants at the current pericenters,
despite having similar semi--major axes and thus binding
energies. These two orbits have significantly smaller eccentricities,
implying higher angular momenta. This strongly suggests that they have
been scattered from their original post--interaction orbits by the
addition of angular momentum near apocenter. For this to work, the
mean gravitational relaxation time should not be much greater than the
nuclear lifetimes $\sim {\rm few}\times 10^6$~yr of these stripped
stars. As a guide we use the estimate
\begin{equation}
t_{\rm relax} = 0.34{\sigma^3\over G^2M\rho\ln \Lambda}
\end{equation}
(Binney \& Tremaine, 1987, eq. (8--71), where $\sigma$ is the
velocity dispersion, $M$ the mass of a star, and $\rho$ the stellar
mass density. Strictly, this formula needs some modification for
the presence of the central black hole, but this should not change
things by large factors.) If stars have been fed into the central
regions for a significant fraction of the Galaxy's lifetime there
could currently be $\sim 10^{-6}\times 10^{10} = 10^4$ remnant stars
(mostly white dwarfs, but with a few neutron stars and black holes)
orbiting within the central 0.05~pc, giving $\rho \sim 10^8\msun{\rm
pc}^{-3}$. The remnants spend almost all their lifetimes near
apoastron, so we take $\sigma$ of order the orbital velocities $\sim
150$~km\ s$^{-1}$ there. This gives $t_{\rm relax} \sim 10^7$~yr. This
supports the idea that the orbits of S1 and S13 have been scattered
once from their original post--interaction forms. We note that these
orbits are not relaxed, and still fairly close to the original ones,
as expected if the number of scatterings is low.

Stellar--mass black holes produced in massive supernovae may
settle into the central parsec by dynamical friction 
(Morris 1993; Miralda-Escude \&  Gould 2000). In doing
so they may scatter the white dwarfs out of the central parsec.
The resulting central density will be similar to that
calculated above (see also Alexander \& Livio 2004), but
the average mass of the stars would be a factor of ten higher,
which would shorten the relaxation time.

\section{Discussion}

We have shown that the S~stars can result from violent stripping of
red giants near pericenter, making them into rather blue objects.
Their subsequent evolution is straightforward. Their nuclear
luminosities are essentially unaffected by the loss of much of the
envelope, so they evolve at constant luminosity until the remaining
envelope mass is burnt. This takes of order a few $\times
10^6$~yr. Their radii shrink and their effective surface temperatures
$T$ increase during this phase.  The constant total luminosity fixes
$AT^4$ ($A=$ area), while the K--band flux varies as $\sim AT$, so
they become fainter in the K--band as $\sim T^{-3}$. This is probably
the reason why all the observed S--stars appear to have very similar
temperatures. A star stripped on the RG branch (i.e. pre-helium
ignition) has a luminosity $\sim 10^2$ L$_\odot$, and with an
effective temperature $T_{\rm eff} \sim 10^4$~K would not be visible
in IR surveys. We expect the S~stars to remain observable for some
fraction of the nuclear timescale $\sim 10^6$~yr.

To maintain a steady state, S~stars must be created at the same rate
$\sim 5 \times 10^{-6}~{\rm yr}^{-1}$ (assuming of there are of order
5 at any moment). This fixes the rate at which stars must be scattered
into near--parabolic orbits with the very small pericenters observed
for the S~stars. The total stellar tidal disruption rate by the
central black hole is $\sim 5 \times 10^{-4}$ yr$^{-1}$, scaling as
$R_\star^{1/4}$ (Wang \& Merritt 2004). Considering the evolutionary
tracks plotted in Fig. 1, we find that about 1 \% of stripped 1--4
M$_\odot$ stars are stripped while on the AGB. Hence the tidal
disruption rate of AGB stars agrees well with the formation rate of
the S~stars in our picture.

The tidal disruption rate derived above implies that gas is injected
into orbit about the black hole at a rate $\sim 2.5\times
10^{-4}\msun~{\rm yr}^{-1}$. This is significantly higher than
estimates of the black hole accretion rate (Baganoff et al
2003). However this is not suprising, as the gas still has to lose
almost all its angular momentum in order to accrete. It is likely that
accretion is a highly time--dependent phenomenon, with the Eddington
limit playing a significant role in determining the growth rate of the
black hole mass.

We note finally that our picture implies the presence of a kind of
``Oort cloud'' of stellar remnants consisting mainly of white dwarfs
around the central black hole. The dynamical evolution of this system
may be very interesting. Moreover the stars in this system must
interact with the orbiting gas discussed in the last paragraph. This
may be the origin of the observed X--ray flares in a similar manner to
that suggested by Nayakshin \& Sunyaev (2003). We might expect the gas
to virialise with the stellar population, implying X--ray temperatures
($\sim 10^8$~K) near the black hole.

In a quite general way, our picture suggests that the stellar
populations near the Galactic Center should show orbital, positional,
and surface characteristics determined by their interaction with the
central black hole. We will return to these questions in future
papers.

\acknowledgements

MBD acknowledges the support of a research fellowship awarded by
the Royal Swedish Academy of Sciences. ARK acknowledges support from PPARC and
a Royal Society Wolfson Research Merit Award.

\begin{figure} 
\plotone{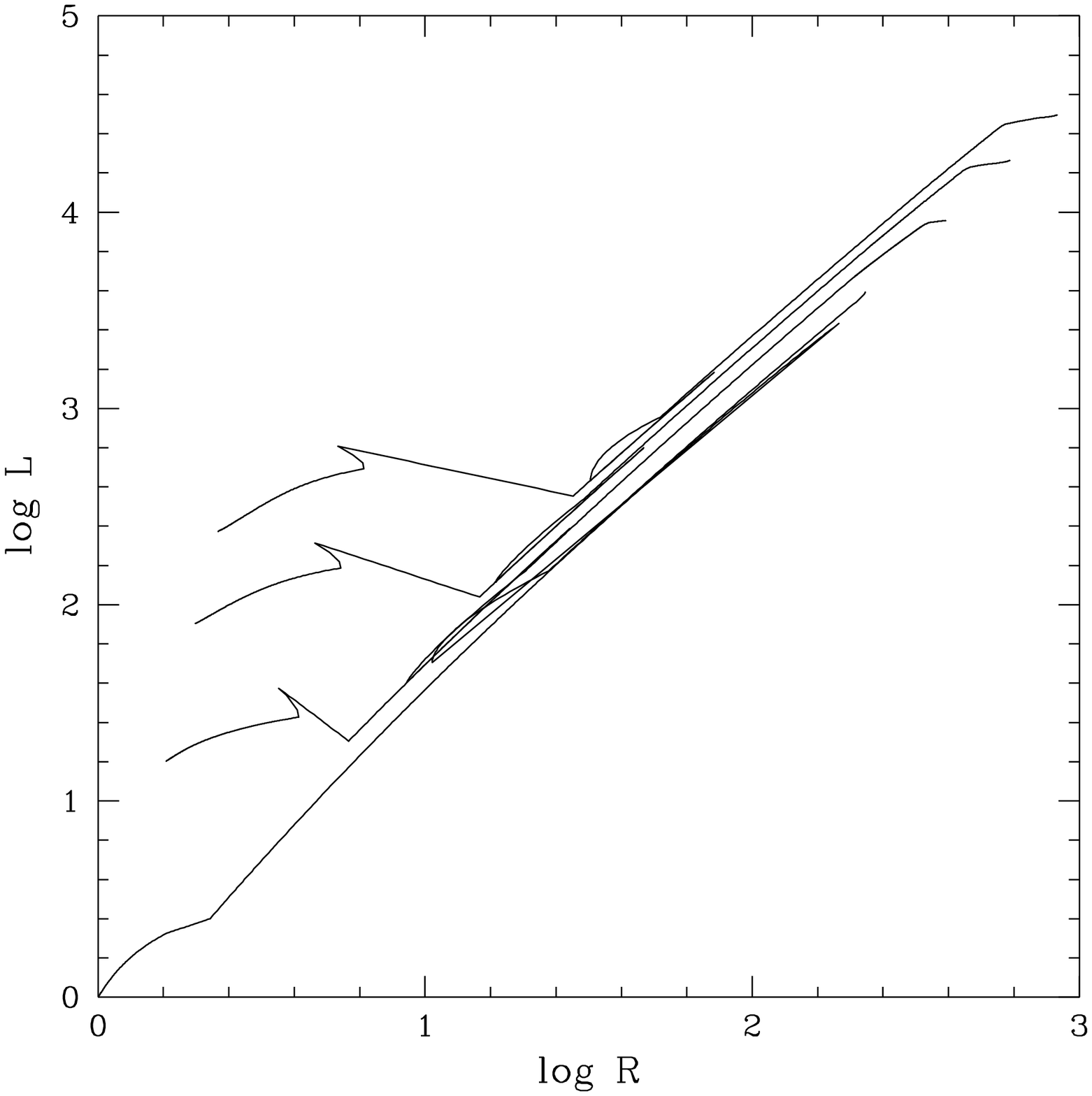}
\caption{Plot of log(luminosity) as a function of log(radius), both
given in solar units, for 1, 2, 3, and 4 M$_\odot$ solar metallicity
stars using the SSE code of Hurley, Pols \& Tout (2000).}
\end{figure}

\end{document}